\documentclass[nonacm, sigconf,10pt]{acmart}
\renewcommand\footnotetextcopyrightpermission[1]{}
\usepackage[english]{babel}
\usepackage{blindtext}

\acmDOI{}

\acmISBN{}

\usepackage{url}
\usepackage{graphicx}
\usepackage{amsmath}
\usepackage{amsthm}
\usepackage{booktabs}
\usepackage{algorithm}
\usepackage[noend]{algpseudocode}
\usepackage{textcomp}
\usepackage{comment}
\usepackage{xcolor}
\usepackage{subfig}
\usepackage{array}
\usepackage{afterpage}
\usepackage{float}
\usepackage{listings}
\usepackage{multirow}
\usepackage{arydshln}
\usepackage{xspace}
\usepackage{enumitem}
\usepackage[htt]{hyphenat}
\usepackage{cleveref}
\usepackage{pifont}

\graphicspath{{figures/}}

\usepackage{cleveref}
\crefformat{chapter}{\S#2#1#3}
\crefformat{section}{\S#2#1#3} 
\crefformat{subsection}{\S#2#1#3}
\crefformat{subsubsection}{\S#2#1#3}

\usepackage{pifont}
\usepackage{xcolor}

\newcounter{wangexpno} 

\newcommand{\tabref}[1]{Tab.~\ref{#1}}
\newcommand{\figref}[1]{Fig.~\ref{#1}}
\usepackage{adjustbox}

\newcommand{\sysname}{CALVO}
\newcommand{\calvo}{\textit{\sysname}\xspace}

\newcommand{\phm}[1]{\vspace{.4em} \noindent\textbf{#1}\hspace{.5em}}
    
\begin{document}
    \title{\sysname: Improve Serving Efficiency for LLM Inferences with Intense Network Demands}

    \author{Weiye Wang}
    \affiliation{%
      \institution{Shanghai Jiao Tong University}
      \country{}
    }

    \author{Chen Chen}
    \authornote{Chen Chen is the corresponding author.}
    \affiliation{%
      \institution{Shanghai Jiao Tong University}
      \country{}
    }

    \author{Junxue Zhang}
    \affiliation{%
      \institution{University of Science and Technology of China}
      \country{}
    }

    \author{Zhusheng Wang}
    \affiliation{%
      \institution{Huawei}
      \country{}
    }

    \author{Hui Yuan}
    \affiliation{%
      \institution{Huawei}
      \country{}
    }

    \author{Zixuan Guan}
    \affiliation{%
      \institution{Huawei}
      \country{}
    }

    \author{Xiaolong Zheng}
    \affiliation{%
      \institution{Huawei}
      \country{}
    }

    \author{Qizhen Weng}
    \affiliation{%
      \institution{Institute of Artificial Intelligence (TeleAI), China Telecom}
      \country{}
    }

    \author{Yin Chen}
    \affiliation{%
      \institution{Institute of Artificial Intelligence (TeleAI), China Telecom}
      \country{}
    }
    \author{Minyi Guo}
    \affiliation{%
      \institution{Shanghai Jiao Tong University}
      \country{}
    }

    \begin{abstract}
        Distributed prefix caching has become a core technique for efficient LLM serving. However, for long-context requests with high cache hit ratios, retrieving reusable KVCache blocks from remote servers has emerged as a new performance bottleneck. Such \emph{network-intensive LLM inference} is expected to become increasingly common as agentic AI workloads continue to grow. However, existing LLM inference engines remain largely compute-centric: they treat KVCache loading as a subordinate phase to GPU execution and often fail to account for its delay explicitly during scheduling. 
        
        We present \calvo, an LLM serving engine that treats KVCache loading as a first-class concern. \calvo decouples KVCache loading and GPU computation into independently managed, asynchronously progressing stages, enabling better utilization of network, PCIe, and computation resources. In addition, \calvo incorporates KVCache loading delay as an explicit component of per-request service cost, leading to more accurate scheduling decisions. Experiments on a real testbed with diverse long-context workloads show that \calvo substantially improves the efficiency of network-intensive LLM inference, achieving up to 61.67\% higher SLO attainment than the baseline.
    \end{abstract}

    \maketitle
	
\section{Introduction}
\label{sec:intro}

Large language models (LLMs) have demonstrated versatile capabilities on various real-world tasks like chatbot conversation and code generation~\cite{wu2025survey,gu2023llm,team2025qwen251m}.
Given the surging user demands, LLM service providers need to efficiently serve the incoming inference requests, for which Time-to-First-Token (TTFT) is a key efficiency metric. 
In particular, confronting the trend of increasing input context length (e.g., for document QA and project-level code completion), \emph{prefix caching}~\cite{qin2025mooncake} is prevalently adopted to optimize TTFT, which, by reusing previously-generated KVCache upon prefix match, leverages storage resources for less computation.

\begin{figure}
    \centering
    \includegraphics[width=0.7\linewidth]{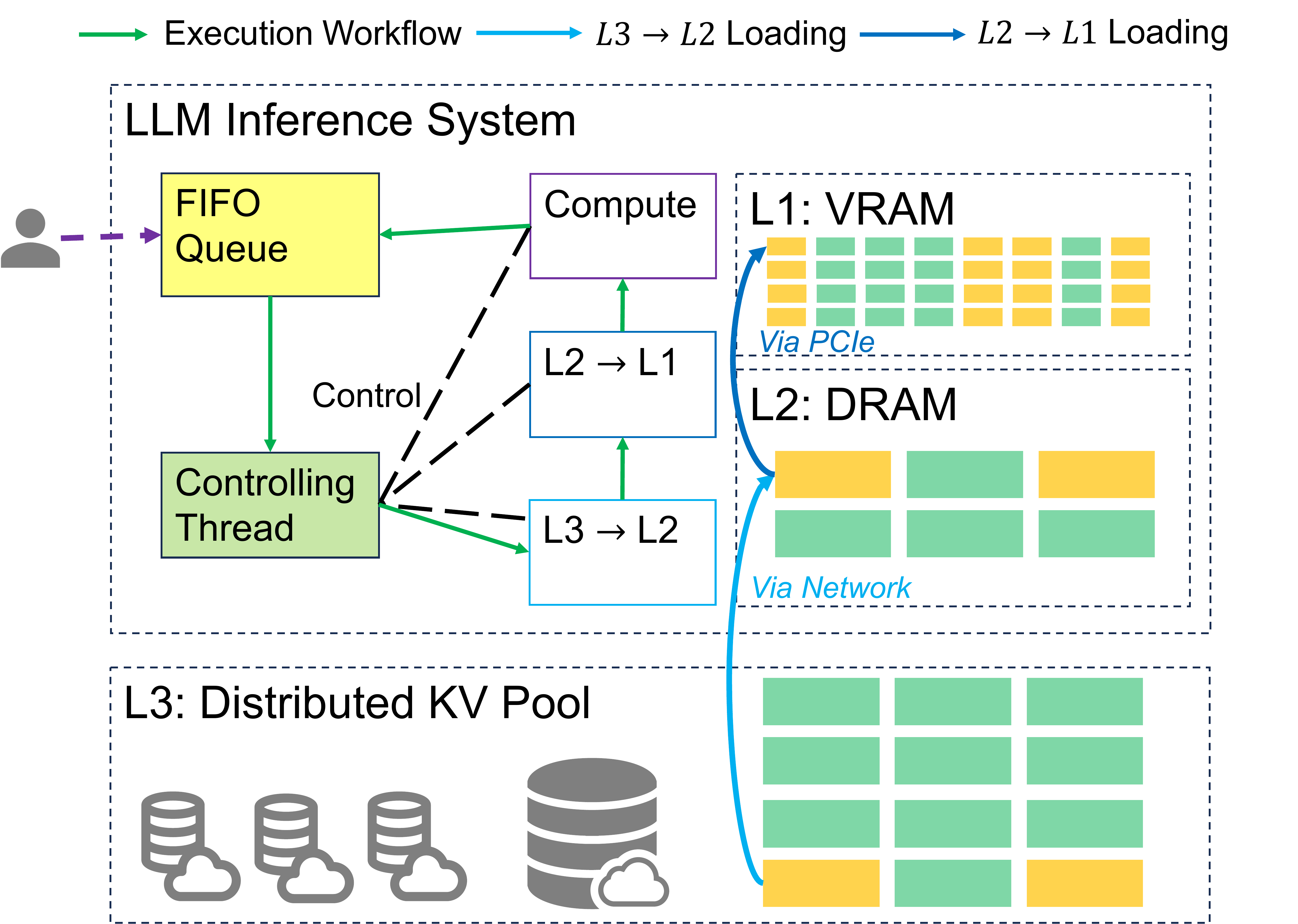}
    \caption{Workflow in a typical LLM inference engine (e.g., vLLM) when integrated with a distributed KVCache pooling framework (e.g., LMCache).} 
    \vspace{-.1in}
    \label{fig:kv-loading-arch}
\end{figure}

To host large-volume KVCache in production platforms, distributed caching is often adopted with the LLM inference engine, rendering it a common demand to conduct inter-node KVCache reuse.
As reported by existing works~\cite{qin2025mooncake}, a single node often fails to provide sufficient memory capacity to attain high cache hit ratio.
To enlarge the available caching space, KVCache storage frameworks such as Mooncake~\cite{qin2025mooncake} and LMCache~\cite{lmcacheweb} combine the DRAM spaces on distributed servers into a shared caching pool. 
As shown in \figref{fig:kv-loading-arch}, before the GPU computation of each inference starts, the LLM engine needs to load the reusable prefix cache from remote memory (L3)---first to its local CPU memory (L2) and then to its GPU HBM (L1).

While inter-node KVCache reuse can improve the cache hit ratio and reduce redundant computation~\cite{qin2025mooncake}, it however introduces non-negligible communication overhead. 
Our empirical measurements show that, when serving a long-context LooGLE dataset~\cite{li2024loogle} with a distributed KVCache pool (with 400 Gbps links), the KVCache loading time may account for over 90\% 
of the TTFT time. 
We call such inference workloads---\emph{which rely heavily on prefix caching and have a communication time comparable to computation}---as \emph{network-intensive LLM inferences}.
With the wide adoption of KVCache pool and the high prompt similarity in popular LLM applications~\cite{lin2024parrot}, network-intensive LLM inferences would be increasingly prevalent in production clusters.

However, when serving \emph{network-intensive} LLM inferences, existing LLM service engines~\cite{kwon2023vllm,zheng2024sglang}---designed for conventional \emph{computation-intensive} inference workloads---often fail to attain high efficiency.
Those frameworks are designed to be \emph{compute-centric}, treating cross-node KVCache loading as a \emph{subordinate step} controlled by the LLM compute scheduler.
This design choice leads to two problems.
First, the centralized, compute-centric serving control incurs low utilization of both compute and network resources. 
Before an inference completes all its stages (i.e., L3-to-L2 loading, L2-to-L1 loading, and GPU computing), no other inferences can simultaneously utilize the temporally idled resources in any stage, causing substantial resource idling. 
Second, compute-centric scheduling control often leads to suboptimal service order. 
When determining the inference service order under contention, existing frameworks rely on simple heuristics like FIFO~\cite{kwon2023vllm} or only consider the computation amount~\cite{du2025prefillonly}, totally ignoring the KVCache loading delay which is actually a significant part of the overall service cost, and this would compromise the overall TTFT performance. 

In this paper, we design \calvo, an optimized LLM engine for efficient serving of network-intensive LLM inferences, guided by the design philosophy to \emph{treat cross-node KVCache loading as a first-class citizen alongside computation}.
To be concrete, there are two key techniques in \calvo.

First, regarding cross-stage service coordination, \calvo decouples control across KVCache loading and computation, so as to fully pipeline the inference serving path. A key challenge is that loading can proceed only after the destination space at the higher storage level has been allocated. To address this challenge, \calvo manages each KVCache loading stage with an independent dispatcher-executor pair, and lets lower-level dispatchers proactively trigger space allocation at higher levels, so that loading at different stages can overlap as soon as data dependencies are satisfied.

Second, regarding inter-inference scheduling order, \calvo incorporates the KVCache loading delay of an inference as part of its overall service cost, which is more accurate and can help to optimize the scheduling performance. 
With system profiling, we depict the service cost as a binary linear function which joins the factors of KVCache loading and computation. 
That cost is then used for working out the best scheduling order---respectively for minimizing the average TTFT and for maximizing SLO attainment.

\calvo is implemented atop vLLM~\cite{kwon2023vllm}, a mainstream LLM serving engine (together with LMCache~\cite{lmcacheweb}, a popular KVCache access interlayer).
We further verify the effectiveness of \calvo by conducting a comprehensive evaluation with diverse benchmarks. 
The results show that, when serving network-intensive LLM inferences, \calvo attains significantly higher efficiency the status-quo practice, with an improvement of 
up to 61.67\% in SLO attainment of TTFT.

\section{Background and Motivation}
\label{sec:background}

\subsection{LLM Inference: The Basics}

\phm{LLM inference: execution procedures and service objectives.}
Large language models (LLMs)~\cite{brown2020language,floridi2020gpt,touvron2023llama} have nowadays been widely adopted in various fields like finance~\cite{li2023large}, arts~\cite{makridis2025impact} and science~\cite{ren2025towards}.
Applying LLM techniques requires LLM inference, which contains two phases: \emph{prefill} and \emph{decode}~\cite{zhong2024distserve}.
The prefill phase processes the entire input prompt 
and constructs the intermediate attention keys and values (known as the \emph{KVCache}), which is commonly deemed computation-intensive.

To LLM service providers, efficient serving competing inference requests from diverse users is crucial.
Specifically, a key service quality metric is the time gap between the request arrival and the generation of the first token, namely \emph{Time-To-First-Token} (TTFT), which critically affects user experience. 
In this paper, we primarily consider accelerating the prefill phase\footnote{
With the wide adoption of PD disaggregation~\cite{zhong2024distserve} as well as the emergence of prefill-only workloads~\cite{du2025prefillonly}, it is increasingly common to independently optimize the TTFT performance for the prefill phase.
}, and focus on optimizing the TTFT performance.

\phm{Prefix caching: a common technique to save computation cost.}
Mainstream LLM today now supports context window of 128K-2M tokens, enabling more long-context tasks like long-document QA~\cite{li2024loogle}.
However, increased context size also introduces substantial computation demands, yielding a quadratic complexity in prefilling computation~\cite{jiang2024minference}. 
To mitigate this, production-grade LLM systems widely adopt prefix caching, which stores the KVCache for previously processed text prefixes~\cite{liu2024cachegen,agarwal2025cachecraft}. 
By reusing KVCache across requests sharing the same prefix, the LLM service system can avoid expensive re-computation during the prefill phase, yielding a much better TTFT performance. For example, as illustrated later in \figref{fig:single_request_loading_breakdwon}, compared with performing pure computation using vLLM, remotely loading the KVCache by vLLM-LMCache can reduce the TTFT by over 88\%.

\subsection{Network-Intensive LLM Inference}

\phm{Prevalence of distributed KVCache storage.}
With the ever-expanding long-context processing demands, the KVCache storage consumption keeps booming at a high rate, rendering the KVCache storage space on a single node easily used up. 
For instance, when serving inference with the Llama-3.1-8B model, the KVCache for merely 10M tokens would consume a storage space of over 1TB.
To accommodate the large-volume KVCache content with relatively high speed (SSD, HDD or cloud storage platforms like S3 is often too slow), modern LLM service providers~\cite{qin2025mooncake} often join the idled memory on different servers into a distributed KVCache pool~\cite{qin2025mooncake,lmcacheweb,hicache_blog}, which is then plugged to the LLM inference engine~\cite{kwon2023vllm,zheng2024sglang}.

As previously shown in \figref{fig:kv-loading-arch}, there are three storage layers for the KVCache content: L1---the GPU VRAM (HBM), L2---the local CPU DRAM, and L3---the remote CPU DRAM. 
When a LLM request arrives, it will wait for being scheduled in a FIFO queue.
Once a request is scheduled into a computation iteration, before the computation starts, the LLM engine needs to load the reusable KVCache content to GPU VRAM.
Specifically, the engine first searches for the longest reusable prefix in the KVCache pool; the spotted KVCache prefix (usually on remote host) would be loaded first from L3 to L2 and then from L2 to L1. 
The inference computation will be triggered once all the KVCache needed by the current execution batch is done. 
After computation, the main control thread will go back to fetch a set of waiting requests to execute in the next iteration.

\begin{table}
\caption{Statistics of representative long-context datasets. ``Avg. Context'' denotes the average number of tokens in the provided context, and ``Avg. Query'' denotes the average token length of the user query.} 
\begin{center}
\begin{footnotesize}
\begin{tabular}{|c|c|c|c|}
\hline
\textbf{} & \textbf{Num. Reqs} & \textbf{Avg. Context} & \textbf{Avg. Query} \\ \hline
\textbf{LooGLE} & $120$ & $28.1K$ & $28$ \\ \hline
\textbf{ICL} & $120$ & $28.3K$ & $61$ \\ \hline
\textbf{Code} & $100$ & $38.3K$ & $209$ \\ \hline
\end{tabular}
\label{tab1:eval_datasets}
\end{footnotesize}
\end{center}
\end{table}

\phm{KVCache loading: an emerging bottleneck for LLM inference.}
In typical LLM applications, an LLM inference input is composed of two parts: a static \emph{application-context} and a dynamic \emph{user-query}.
The KVCache content of the shared context is usually reusable across different requests, with runtime computation only needed for processing the user query.
In particular, with the in-depth adoption of LLM techniques, the static contexts in many applications are increasingly complex and lengthy. 
As shown later in \tabref{tab1:eval_datasets}, scenarios such as long document QA and project-level code completion task involve extremely long contexts (e.g., an average of 28.1K tokens in LooGLE~\cite{li2024loogle}), whereas the corresponding user queries are relatively short (e.g., only 28 tokens on average).  
This highlights an increasingly common pattern in the era of agentic AI era~\cite{lin2024parrot,ren2025towards}: \emph{long context, short query}.

\begin{figure}[t]
    \centering
    \vspace{-.1in}
    \includegraphics[width=\columnwidth]{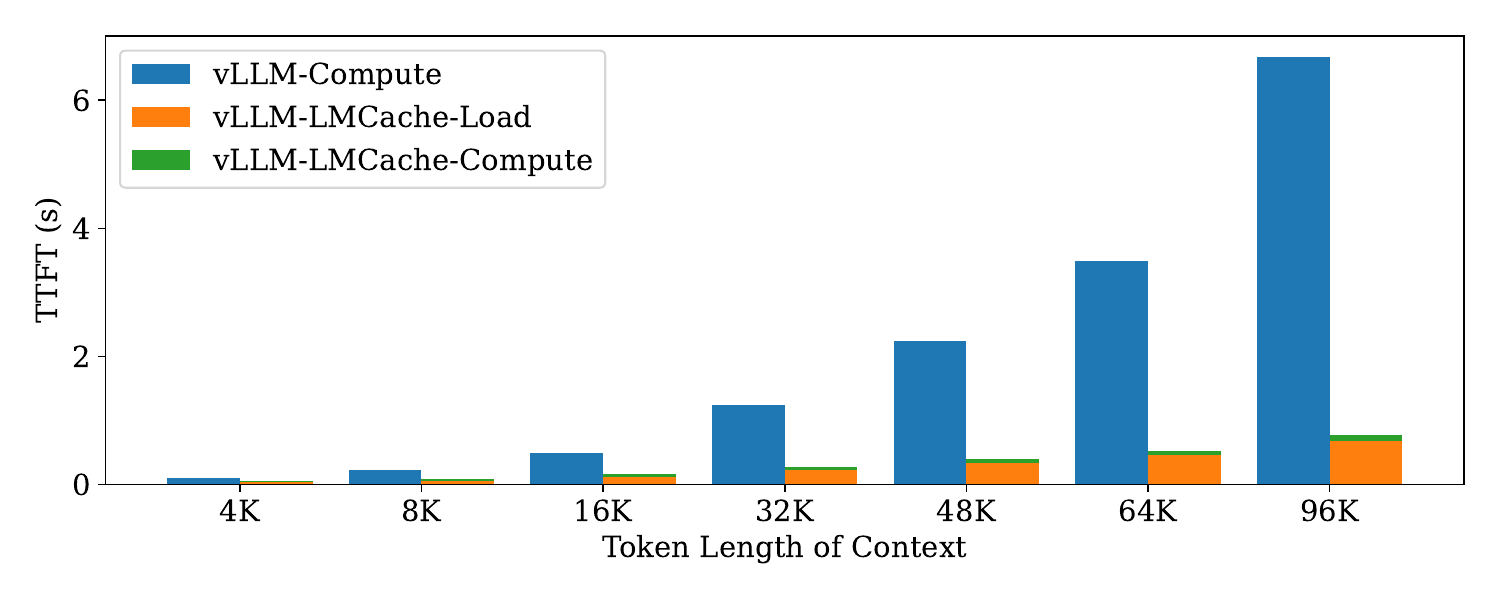}
        \caption{Breakdown of TTFT when a Llama-3.1-8B model serves requests with varying context-token length (to load from a remote server) yet fixed query-token length. From the figure we learn that KVCache reusing can effectively reduce TTFT, yet KVCache loading now becomes a major bottleneck.}
        \vspace{-.1in}
    \label{fig:single_request_loading_breakdwon}
\end{figure}

Since KVCache reusing essentially trades storage for computation efficiency, a larger context-to-query ratio (i.e., a higher cache hit ratio) would shift the bottleneck from KVCache processing (computation) to \emph{KVCache loading} (communication). 
\figref{fig:single_request_loading_breakdwon} shows the TTFT breakdown when serving a single request with varying context lengths yet a fixed query length (1000); the KVCache is loaded from a remote server following the setup in \S\ref{sec:setup}.
It clearly demonstrates that, while KVCache reusing can help to reduce the overall TTFT, the ratio of KVCache loading in TTFT keeps increasing, becoming an remarkable performance bottleneck.

For such LLM workloads with a high cache reuse ratio, since network transmission time is a key building block of their KVCache loading delay, we call them as \emph{network-intensive LLM inference}.
As LLM applications become more complex, it can be expected that network-intensive LLM inference would be a mainstream LLM workload type in the future.
Therefore, in this paper we focus on efficiency (TTFT) optimizations of network-intensive LLM inferences.

\subsection{Inefficiencies of Existing Practices}
\label{sec:motivation}

While network-intensive LLM inferences are increasingly popular, existing LLM serving engines~\cite{kwon2023vllm,zheng2024sglang} 
still follow the legacy design for 
conventional computation-dominate LLM workloads.
Our study shows that their \emph{compute-centric} design renders themselves high inefficient for network-intensive inferences, and there are two reasons for that.

\subsubsection{Low resource utilization due to centralized, compute-centric control of the entire serving workflow.}

\begin{figure}
    \begin{minipage}[t]{0.55\columnwidth}
        \centering
        \includegraphics[width=\textwidth]{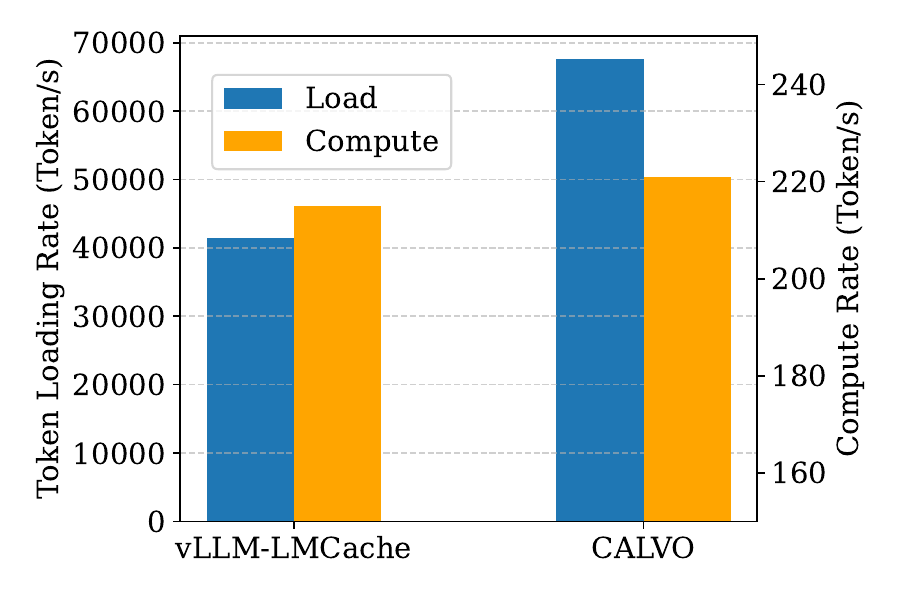}
        \caption{Per-stage processing throughput in vLLM-LMCache is suboptimal (the throughput of our later proposed \calvo system is shown for comparison). }
        \label{mot:bandwidth_utilization}
    \end{minipage}
    \hfill
    \begin{minipage}[t]{0.40\columnwidth}
        \centering
        \includegraphics[width=\textwidth]{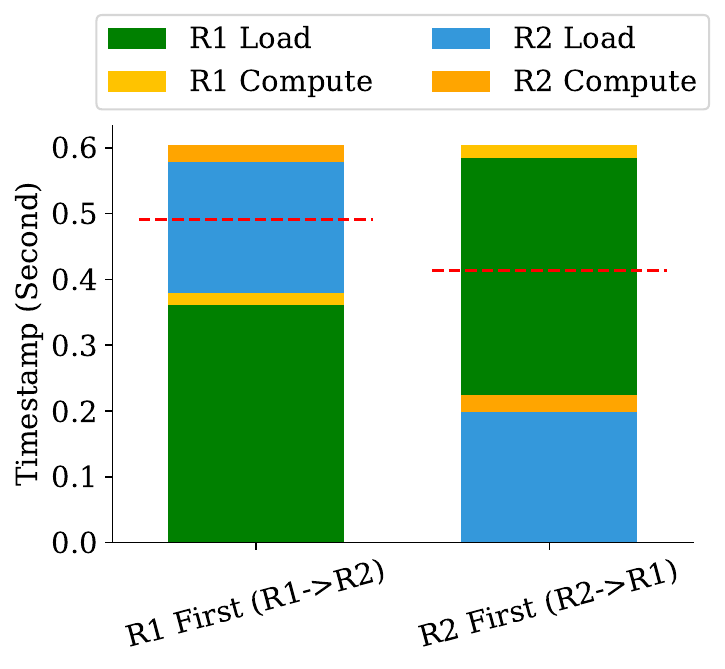}
        \caption{Given two network-intensive requests, FIFO or compute-based SJF may prolong the average TTFT.} 
        \label{mot:loading-aware-scheuduling}
    \end{minipage}
\end{figure}
Different KVCache processing stages in \figref{fig:kv-loading-arch} require different resource types: 
network bandwidth for L3$\to$L2, PCIe bandwidth for L2$\to$L1, and GPU processors for computation.
However, the current LLM engines exemplified by vLLM adopts a centralized (single-threaded) control over the entire workflow: each stage (including the computation stage) is triggered passively as instructed by the main thread. 
Such a centralized control is reasonable for conventional LLM workloads where there is no runtime KVCache loading, but no longer so for the emerging network-intensive LLM inferences:
for which a centralized control is straightforward.
when the KVCache blocks of one inference request step to the next stage, the idled resources cannot be immediately used to serve other pending requests, leading to poor resource utilization.

Our measurements confirm this inefficiency.
Specifically, we serve the LooGLE dataset~\cite{li2024loogle} over the Llama-3.1-8B model, with a workload intensity of 1.2 queries per second. We measure per-stage processing throughput as the peak average throughput within any 20-second interval. In \figref{mot:bandwidth_utilization} we compare such per-stage processing throughput under vLLM-LMCache with that under our optimized \calvo engine (elaborated later in \S\ref{subsec:inter-stage-workflow}).
The comparison results clearly demonstrate the resource utilization of current LLM serving frameworks is far from optimal. 

\subsubsection{Low scheduling quality due to blindness to KVCache loading costs.} 
\label{subsec:mot_scheduling_necessity}

On production LLM service backends, a massive number of LLM inference submitted from different users may compete for the limited resources;
under such resource contention, the scheduling order critically affects the overall performance---in terms of the average TTFT value or the overall SLO attainment (in cases where a maximum allowable TTFT is associated for each request~\cite{liu_andes_2024}). 
While KVCache loading has become a main performance bottleneck for network-intensive LLM inference, existing LLM engines are blind to such costs in scheduling.
Such blindness compromises the LLM scheduling quality.

To illustrate the impact, we profile the loading and computation time of requests from the loogle dataset~\cite{zou2025manyshot} with vLLM-LMCache. 
Consider two sampled requests: R1 (arriving earlier)---requiring 0.361 s to load and 0.019 s to compute, and R2---requiring 0.199 s to load and 0.025 s to compute. 
A scheduler that enforces naive FIFO or enforces SJF based on only the computation time~\cite{du2025prefillonly} would prioritize R1, resulting in an average TTFT of 0.49 s.
In contrast, another scheduler applying SJF with awareness to the KVCache loading time (prioritizing R2) can reduce the average TTFT to 0.41 s.

\phm{Summary.} 
In summary, the compute-centric nature of existing LLM engines hurts the serving efficiency of network-intensive LLM inferences; to improve, we need to optimize existing LLM service engines in both inter-stage service coordination and inter-request queue management. 

\section{Solution}
\label{sec:solution}

\begin{figure}
    \centering
    \includegraphics[width=0.85\linewidth]{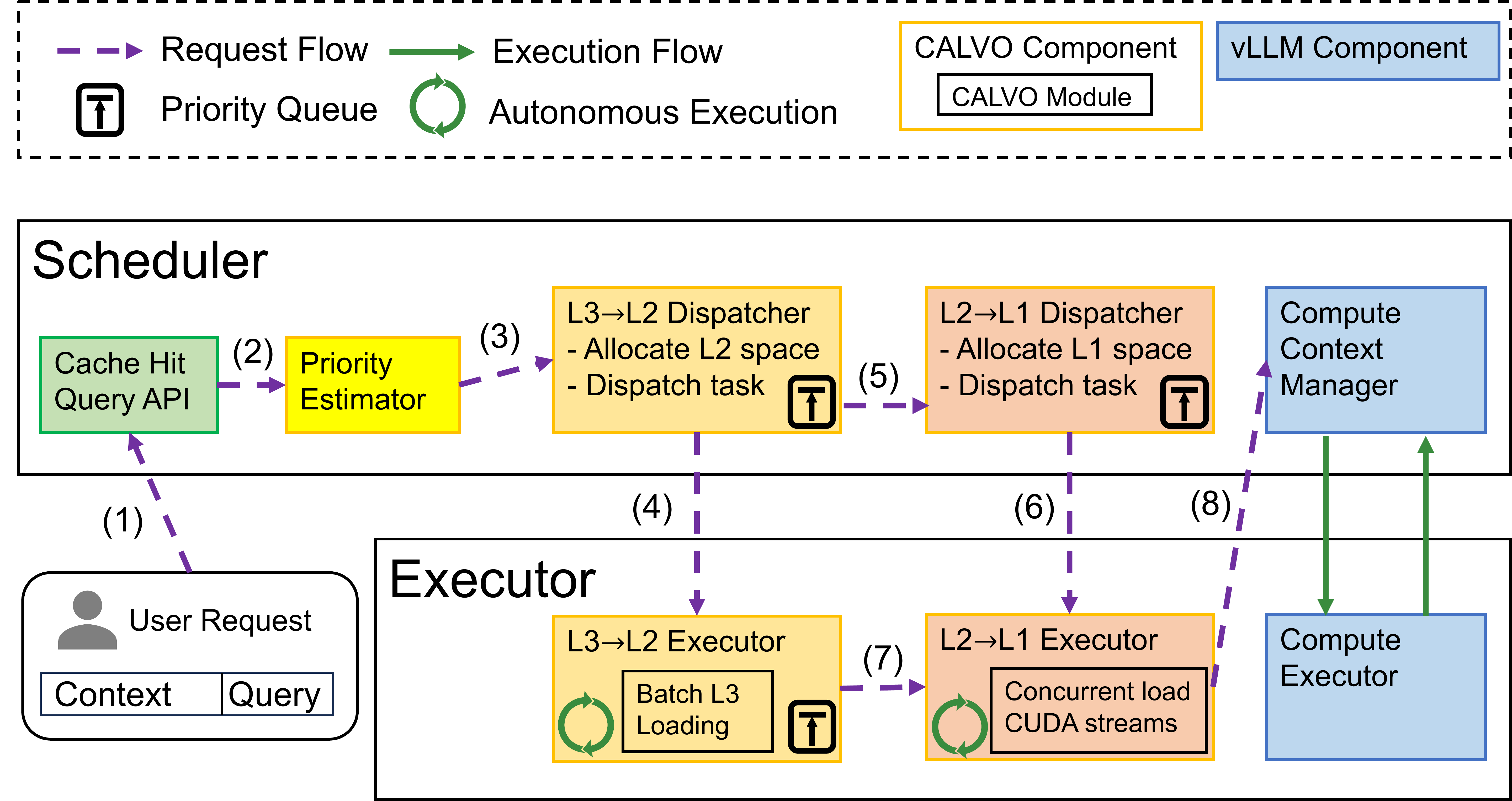}
    \caption{Workflow of \calvo.} 
    \label{fig:calvo_workflow}
\end{figure}

In this section, we present 
\calvo
, an optimized LLM engine that---by treating cross-network KVCache loading as a \emph{first-class citizen (stage) equally important to prefill-computation}---can efficiently serve network-intensive LLM inferences. 
As shown in \figref{fig:calvo_workflow}, \calvo is built atop vLLM; its workflow is driven by a priority estimator and multiple loading dispatcher/executors. 
By adding the autonomous dispatchers/executors, \calvo enhances the resource utilization in each stage; meanwhile, under the new priority estimator, \calvo improves the overall scheduling performance with awareness to the KVCache loading costs. 
Next we elaborate the optimizations respectively on the two aspects. 

\subsection{Inter-stage Workflow Optimization}
\label{subsec:inter-stage-workflow}

Recall that our previous study shows that, in existing LLM engines, the centralized control of the KVCache processing workflow (across all the stages of L3 $\to$ L2 loading, L2 $\to$ L1 loading, and GPU computation) compromises the overall resource utilization. 
To maximize the resource utilization on each stage under the dataflow dependency, our insight is to equally grant each stage with the autonomy to act independently---similar to the classical \emph{dynamic scheduling} mechanism in computer architecture design~\cite{Shen1995Microarchitecture}.

To that end, in \calvo we let each KVCache loading stage managed by an independent dispatcher-executor pair. 
As depicted in \figref{fig:calvo_workflow}, for each loading stage, each dispatcher 
keeps driving its executor to conduct KVCache loading as long as \emph{(1) some lower-level KVCache blocks of the highest-priority request are ready to pick up}, and \emph{(2) the destination storage spaces of those KVCache blocks are allocated}. 
For the compute stage, when all KV-cache blocks required by a request finally become resident in L1 space, it immediately launches the prefilling process, avoiding the compute wastage caused by waiting for loading completion in the critical path. 

Meanwhile, for smooth cross-stage negotiation, each time a stage completes loading a KVCache block, it would issue a signal to the upper-level stage executor, so as to enable fine-grained loading overlapping.
In particular, to let KVCache loading be conducted autonomously, L1 space allocation should not be triggered by the compute stage in a reactive manner, which would block KVCache preloading; instead, in \calvo we let the lower-level stage dispatcher proactively\footnote{
While such proactive space allocation may take additional GPU memory space, since the prefill phase is computation-bound, trading additional memory consumption for faster execution is usually acceptable. In the extreme case where the available space in GPU memory in not enough, this method simply degrades to the reactive allocation method.
} trigger space allocation at the higher storage level. 
Similarly, when the L3$\to$L2 dispatcher issues a data transfer task to its executor, it will at the same time submit a GPU memory allocation request to the L2$\to$L1 dispatcher, so as to proactively reserve GPU space for pipelined L2$\to $L1 loading.

\begin{figure}[t]
    \centering
    \vspace{-.15in}
    \includegraphics[width=0.8\columnwidth]{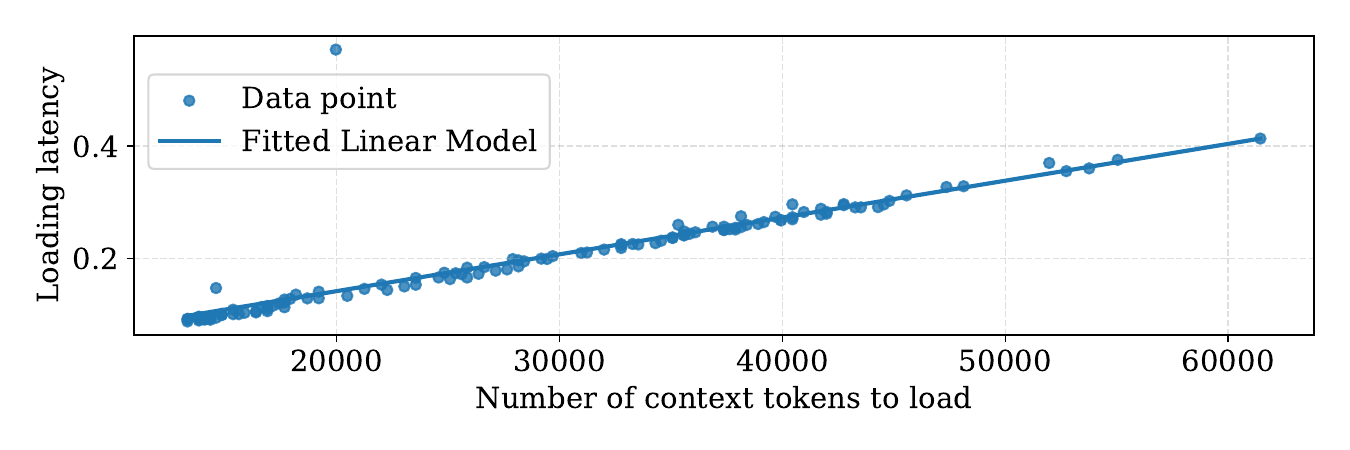}
    \caption{Relationship between loading latency and the number of context tokens to load, when serving Llama-3.1-8B model, following the setup in \S\ref{sec:setup}.}
    \vspace{-.15in}
    \label{fig:load_time_token_relationship}
\end{figure}

\subsection{Inter-request Scheduling Optimization}

To optimize the overall scheduling performance, a prerequisite is to precisely predict the service cost of each LLM inference request; based on the discussion in \S\ref{subsec:mot_scheduling_necessity}, in \calvo we incorporate the KVCache loading delay as \emph{an independent component of the overall cost} when serving a network-intensive LLM inference.
Specifically, 
with offline system profiling, we fit out a \emph{binary linear function} as the performance model, which respectively captures (1) $T_{\text{load}}$---loading latency as a function of the number of context tokens (which is linear as measured by \figref{fig:load_time_token_relationship}), and (2) $T_{\text{comp}}$---computation latency as a function of the number of query tokens.

We then leverage the modeled prefill cost to optimize the scheduling policy. 
For different efficiency objective forms, this translates to different scheduling algorithms. 

(1) To optimize the average TTFT of competing inference requests, we apply the Shortest-Job-First (SJF) principle with the estimated prefill cost, which is known to be optimal~\cite{Cheng1985}. 

(2) To optimize the overall SLO attainment (in cases where each request is associated as a TTFT deadline at submission time), we apply the Least-Slack-Time-First (LSTF) principle, which is optimal for SLO attainment ratio~\cite{scheduling1973}.
For each incoming request, LSTF computes its \emph{slack time} by $\text{LST} = \text{DDL} - T_{\text{load}} - T_{\text{comp}}$. 
The computed per-request LST serves as the priority of the corresponding request---with a smaller LST value indicating a higher urgency.

\section{Evaluation}
\label{sec:eval}

\subsection{Experimental Setup}
\label{sec:setup}

\begin{figure}[t]
    \centering
    \vspace{-.1in}
    \includegraphics[width=0.8\columnwidth]{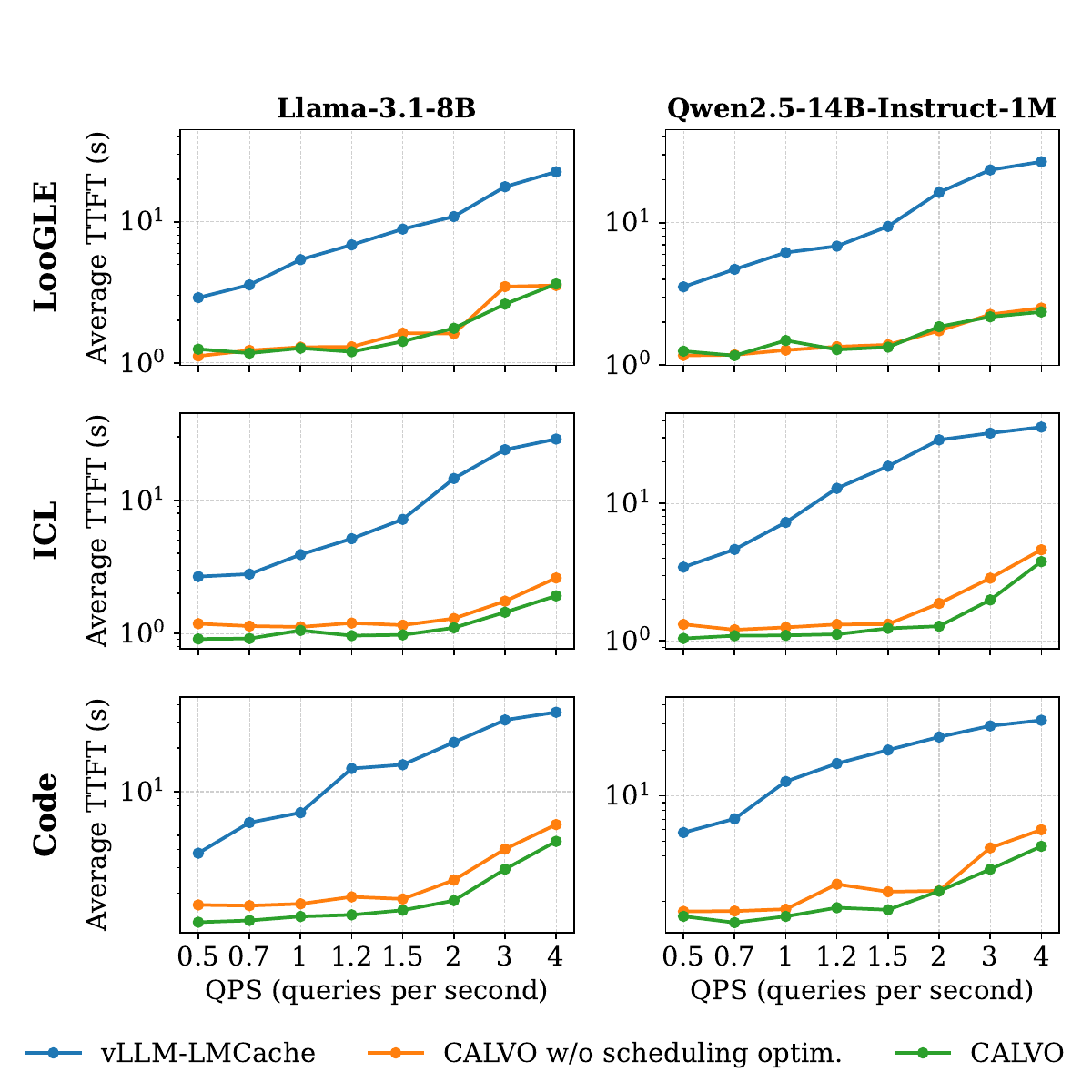}
    \caption{Performance on Average TTFT Latency.}
    \label{fig:eval:latency_by_qps}
\end{figure}
\begin{figure}[t]
    \centering
    \vspace{-.1in}
    \includegraphics[width=0.8\columnwidth]{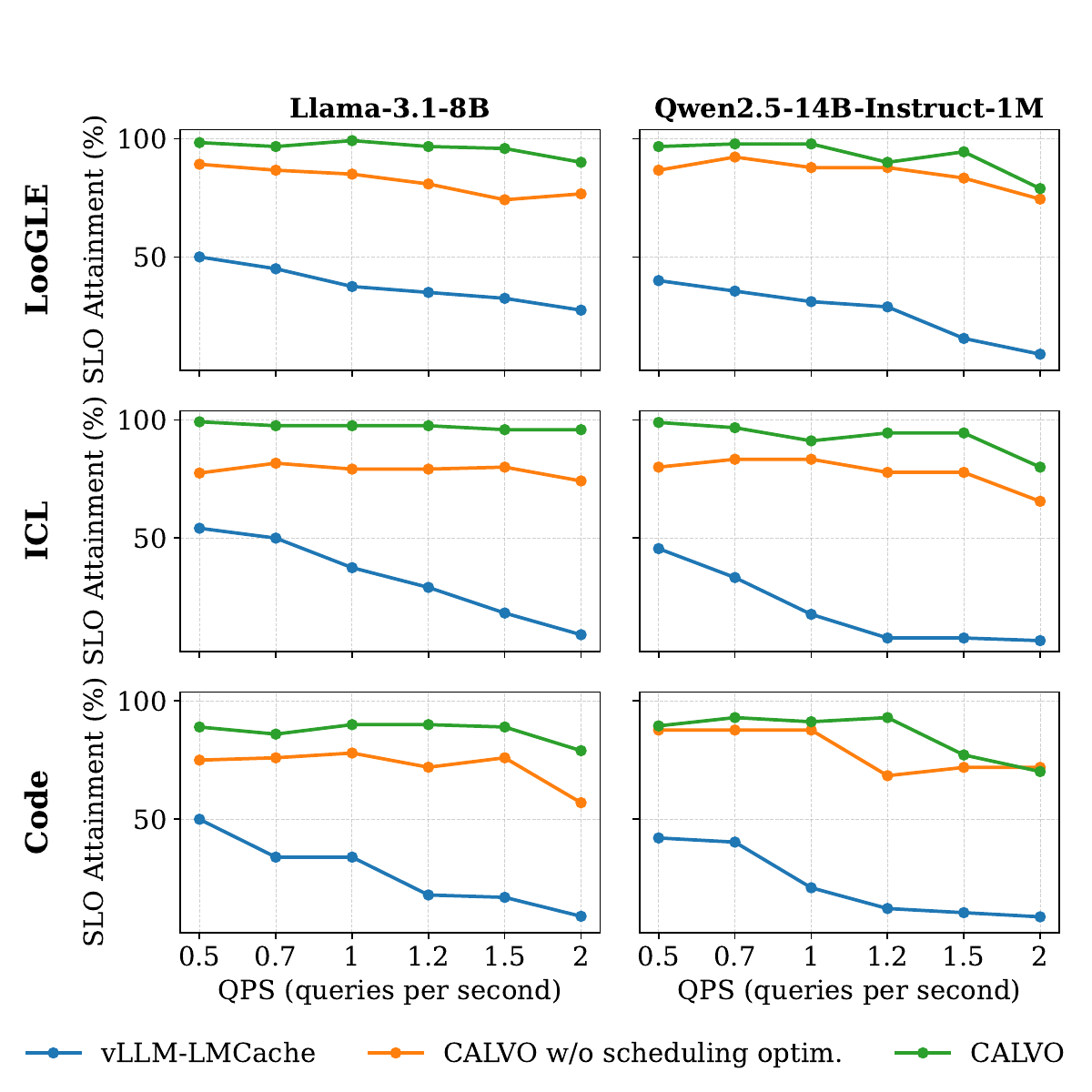}
    \caption{Performance on TTFT SLO Attainment.}
    \label{fig:eval:slo_attainment_by_qps}
\end{figure}

\phm{Implementation.}
We implement\footnote{For fair comparison, we identified and optimized two performance issues in vLLM-LMCache, which are orthogonal to our contributions. First, to ensure LMCache fully utilize loading bandwidth, we use Mooncake's \texttt{batch\_get\_into} interface for L3$\rightarrow$L2 transfers, and we issue multiple L2$\rightarrow$L1 load kernels on different CUDA streams to better utilize PCIe bandwidth. Second, we restrict vLLM-LMCache to prefill only one request at a time, preventing incorrect batching decisions by vLLM default scheduler.}
\calvo on top of vLLM (v0.9.1)~\cite{kwon2023vllm} and LMCache (v0.3.1) ~\cite{lmcacheweb} with 3.3K lines of code, and we use Mooncake Store~\cite{qin2025mooncake} as the L3 KVCache storage backend; all those frameworks are representative in production environments.
To enable direct communication between loading dispatchers and loading executors without interrupting the main computation loop of vLLM, we use the ZeroMQ library ~\cite{zeromqwebpage} to implement inter-process communication between worker process and scheduler process. 
To make the vLLM scheduler agnostic to KV-cache loading, we override the \texttt{add\_request()} function of the vLLM scheduler to intercept inference requests, preventing them from being enqueued in vLLM's FIFO computation scheduler queue.

\phm{Hardware platform.} 
The evaluation is conducted on one GPU node equipped with an 
80 GB GPU and 128 GB CPU DRAM. 
The remote CPU DRAM pool consists of a CPU node with 512 GB DRAM; the two nodes are interconnected via RDMA with a 400 Gbps bandwidth link, all representative in modern GPU clusters.

\phm{Workloads.} 
We utilize two popular open-source LLMs with long context capabilities: Llama-3.1-8B-Instruct~\cite{modelllama3.18b} and Qwen2.5-14B-Instruct-1M~\cite{team2025qwen251m}.
Meanwhile, we construct the test suite by sampling contexts and corresponding queries from the three long context datasets:
\textit{LooGLE}~\cite{li2024loogle} includes long documents from diverse sources such as arXiv, Wikipedia;
\textit{ICL}~\cite{zou2025manyshot} features many-shot In-Context Learning tasks from diverse domains such as classification, summarization, reasoning, and translation;
\textit{Code}~\cite{bogomolov2024long} covers varying coding tasks for LLM, and we use its project level code completion tasks for our evaluation. 
The statistics of these datasets are described in \tabref{tab1:eval_datasets}.
Besides, since request arrival intervals are not available in these datasets, we simulate them following the Poisson distribution with different query intensities.

\subsection{End-to-end Performance}

\phm{Average TTFT.} We first evaluate the end-to-end performance of \calvo under diverse QPS rates. 
As shown in \figref{fig:eval:latency_by_qps}, regarding the {average TTFT} performance, \calvo substantially outperforms the default baseline (vLLM-LMCache) as well as its variant without scheduling optimization (Use FIFO scheduling policy). 
For example, for the ICL dataset, when QPS is 1.2, \calvo can reduce the average TTFT by over 81.3\%. 
Moreover, by comparing \calvo with its FIFO variant, we learn that scheduling optimization in \calvo is indeed indispensable for optimizing the overall performance efficiency.

\phm{SLO Attainment.} To check {SLO attainment} performance, we further assign each request with a TTFT SLO, by scaling its TTFT measured under interference-free conditions with a factor uniformly sampled from $\{2\times, 4\times, 8 \times \}$, which is similar to prior works~\cite{gu2023elasticflow,narayanan2023heterogeneityaware}.
The results in \figref{fig:eval:slo_attainment_by_qps} confirm the performance superiority of \calvo: in each case, the SLO attainment under \calvo is consistently better than the baselines. 
Specifically, when the QPS is 1.2, the SLO attainment of \calvo is 61.67\% higher than the vLLM-LMCache baseline.

\subsection{Micro-benchmark Analysis}

\begin{figure}[t]
    \centering
    \begin{minipage}[t]{0.48\columnwidth}
        \centering
        \includegraphics[width=0.9\textwidth]{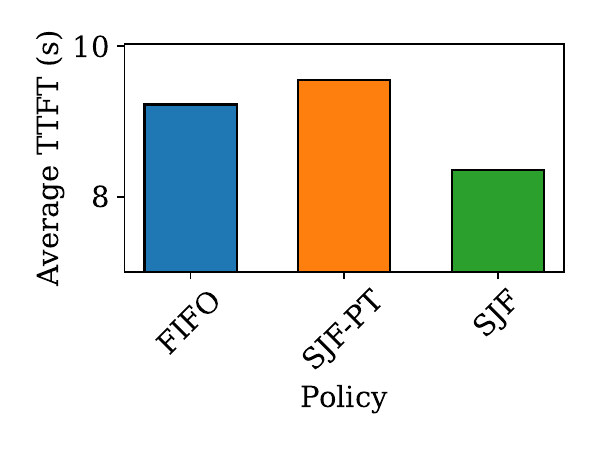}
        \vspace{-.15in}
        \caption{Comparison of different scheduling policies in Average TTFT.} 
        \label{fig:eval:ablation_scheudling_pocliy_avg_ttft}
    \end{minipage}
    \hfill
    \begin{minipage}[t]{0.48\columnwidth}
        \centering
        \includegraphics[width=0.9\textwidth]{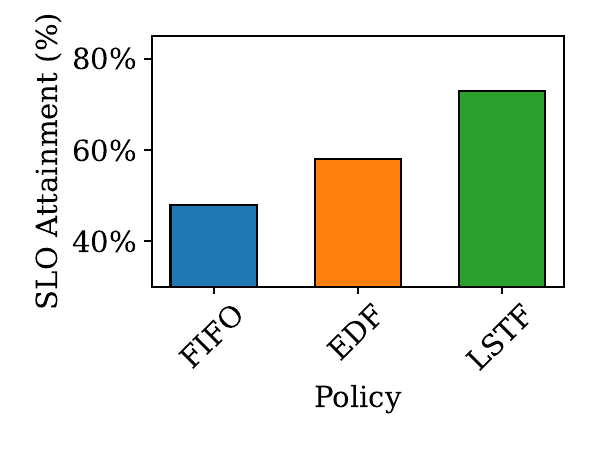}
        \vspace{-.15in}
        \caption{Comparison of different scheduling policies in SLO Attainment.} 
        \label{fig:eval:ablation_scheudling_pocliy_slo}
    \end{minipage}
\end{figure}

\phm{Superiority of binary cost modeling.}
In inference scheduling, another choice of cost modeling is to simply use the prefill token number (we call the resultant scheduling method as \emph{SJF-PT}).
We thus compare \calvo (with the SJF policy integrated) against SJF-PT in \figref{fig:eval:ablation_scheudling_pocliy_avg_ttft}, where each inference request is assigned a cache hit ratio randomly sampled from 25\%, 50\%, 75\% and 100\%. 
\figref{fig:eval:ablation_scheudling_pocliy_avg_ttft} suggests that, when not using the binary linear function for cost modeling, the estimated scheduling priority would be highly inaccurate, rendering the TTFT even worse than FIFO.

\phm{Superiority of LSTF.} 
We also compare \calvo (with the LSTF policy integrated) with the Earliest Deadline First (EDF) policy, the latter not relying on the request service cost.
As shown in \figref{fig:eval:ablation_scheudling_pocliy_slo}, with the concrete knowledge of inference service cost, LSTF can make higher-quality scheduling decisions than the static EDF policy ($73\%$ versus $58\%$).
This confirms the need to leverage the information of KVCache loading cost in scheduling. 

\begin{figure}[t]
    \centering
   
\vspace{-.15in}

    \includegraphics[width=0.8\columnwidth]{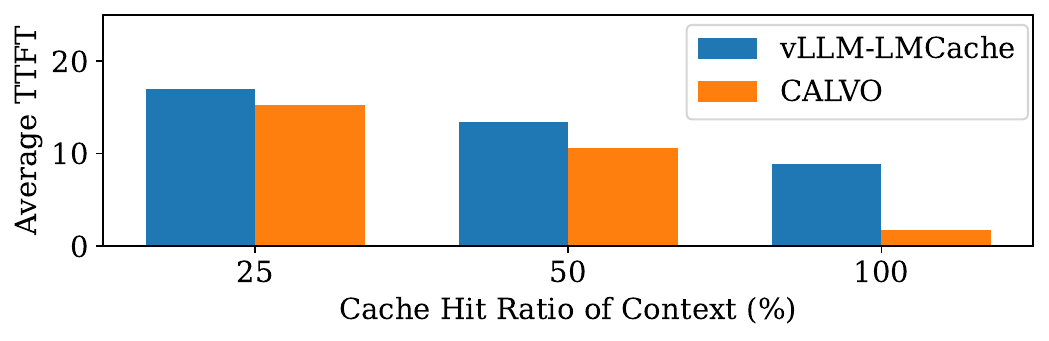}
    \vspace{-.15in}
    \caption{Average TTFT under diverse cache hit rates.} 
    \label{fig:eval:calvo_varying_cache_hit_ratio}

\end{figure}

\phm{Sensitivity study on cache hit ratio.} We manually set the cache hit ratio respectively to 25\%, 50\%, 75\% and 100\%, and \figref{fig:eval:calvo_varying_cache_hit_ratio} shows the resultant \calvo performance.
According to \figref{fig:eval:calvo_varying_cache_hit_ratio}, the average TTFT performance is monotonously improving under \calvo as the cache hit ratio increases, suggesting that it is highly desirable to apply \calvo in the coming agentic AI era with high KVCache reusing ratio~\cite{qin2025mooncake}.

    
\section{Conclusion and Future Work}
\label{sec:conlucsion}

In this paper, we design \calvo, an optimized LLM engine for efficient serving of network-intensive LLM inference. 
\calvo treats cross-server KVCache loading as a first-class stage to serve, granting autonomy to each stage for higher resource utilization, and also incorporating KVCache loading delay as independent cost factor for better scheduling decisions.  
Evaluations on long-context datasets show that \calvo can effectively enhance the serving efficiency of network-intensive LLM inferences in average TTFT and SLO attainment. 

In the future, we plan to extend \calvo to accelerate agentic AI workflows, which requires the correlated KVCache loading tasks be scheduled in a coordinated manner. Meanwhile, we also plan to apply the \emph{network-as-a-first-class-citizen} philosophy into production inference system like Mooncake, so as to enhance the request routing performance as well as to mitigate the network collisions.

    \bibliographystyle{ACM-Reference-Format}
    \bibliography{main}
\end{document}